\documentclass[referee, pdflatex,sn-nature]{sn-jnl}

\usepackage{graphicx}%
\usepackage{multirow}%
\usepackage{amsmath,amssymb,amsfonts}%
\usepackage{amsthm}%
\usepackage{mathrsfs}%
\usepackage[title]{appendix}%
\usepackage{xcolor}%
\usepackage{textcomp}%
\usepackage{manyfoot}%
\usepackage{booktabs}%
\usepackage{algorithm}%
\usepackage{algorithmicx}%
\usepackage{algpseudocode}%
\usepackage{listings}%

\theoremstyle{thmstyleone}%
\theoremstyle{thmstyletwo}%

\theoremstyle{thmstylethree}%

\begin{document}

\title[Lense-Thirring precessing magnetar engine drives a superluminous supernova]{\bf Lense-Thirring precessing magnetar engine drives a superluminous supernova}


\author*[1,2]{\fnm{Joseph R.} \sur{Farah}}\email{jfarah@lco.global}

\author[3]{\fnm{Logan J.} \sur{Prust}}\email{ljprust@kitp.ucsb.edu}
\equalcont{These authors contributed equally to this work.}

\author[1,2]{\fnm{D. Andrew} \sur{Howell}}\email{dahowell@lco.global}
\equalcont{These authors contributed equally to this work.}

\author[1,3]{\fnm{Yuan Qi} \sur{Ni}}\email{cni@lco.global}
\equalcont{These authors contributed equally to this work.}

\author[1]{\fnm{Curtis} \sur{McCully}}\email{cmccully@lco.global}

\author[1,2]{\fnm{Moira} \sur{Andrews}}\email{mandrews@lco.global}

\author[4,5]{\fnm{Harsh} \sur{Kumar}}\email{harsh.kumar@cfa.harvard.edu}

\author[4,5,6]{\fnm{Daichi} \sur{Hiramatsu}}\email{daichi.hiramatsu@cfa.harvard.edu}

\author[9,4]{\fnm{Sebastian} \sur{Gomez}}\email{sebastian.gomez@austin.utexas.edu}

\author[1,2]{\fnm{Kathryn} \sur{Wynn}}\email{kwynn@lco.global}

\author[8]{\fnm{Alexei V.}
\sur{Filippenko}}\email{afilippenko@berkeley.edu}

\author[7]{\fnm{K. Azalee} \sur{Bostroem}}\email{bostroem@arizona.edu}

\author[4,5]{\fnm{Edo}
\sur{Berger}}\email{eberger@cfa.harvard.edu}

\author[4,5]{\fnm{Peter}
\sur{Blanchard}}\email{peter.blanchard@cfa.harvard.edu}

\affil[1]{\orgdiv{Las Cumbres Observatory}, \orgaddress{\street{6740 Cortona Drive, Suite 102}, \city{Goleta}, \postcode{93117-5575}, \state{CA}, \country{USA}}}

\affil[2]{\orgdiv{Department of Physics}, \orgname{University of California}, \orgaddress{ \city{Santa Barbara}, \postcode{93106-9350}, \state{CA}, \country{USA}}}

\affil[3]{\orgname{Kavli Institute for Theoretical Physics}, \orgaddress{ \city{Santa Barbara}, \postcode{93106}, \state{CA}, \country{USA}}}

\affil[4]{\orgname{Center for Astrophysics $\vert$ Harvard-Smithsonian}, \orgaddress{ \city{60 Garden Street, Cambridge}, \postcode{02138}, \state{MA}, \country{USA}}}

\affil[5]{\orgdiv{The NSF AI Institute for Artificial Intelligence and Fundamental Interactions}, \orgaddress{\country{USA}}}

\affil[6]{\orgdiv{Department of Astronomy}, \orgname{University of Florida}, \orgaddress{211 Bryant Space Science Center, \city{Gainesville}, \postcode{32611-2055}, \state{FL}, \country{USA}}}

\affil[7]{\orgname{Steward Observatory, University of Arizona}, \orgaddress{933 North Cherry Avenue, Tucson, \postcode{85721-0065}, \state{AZ}, \country{USA}}}

\affil[8]{\orgdiv{Department of Astronomy}, \orgname{University of California}, \orgaddress{ \city{Berkeley}, \postcode{94720-3411}, \state{CA}, \country{USA}}}

\affil[9]{\orgdiv{Department of Astronomy}, \orgname{The University of Texas at Austin}, \orgaddress{ 2515 Speedway, Stop C1400, \city{Austin}, \postcode{78712}, \state{TX}, \country{USA}}}


\abstract{
Type I superluminous supernovae (SLSNe-I) are at least an order of magnitude brighter than standard supernovae, with the internal power source for their luminosity still unknown \citep{Gal-Yam2017a,Moriya2019,Quimby2019}.
The central engines of SLSNe-I are hypothesized to be magnetars \citep{Kasen2010,Woosley2010}, but the 
majority of SLSNe-I light curves have multiple bumps or peaks that are unexplained by the standard magnetar model \citep{Lunnan2018,Hosseinzadeh2022}. Existing explanations for the bumps either modulate the central engine luminosity or invoke interactions with material in the circumstellar environment.
Systematic surveys of the limited sample of SLSNe-I light curves find no compelling evidence favoring either scenario \citep{Chatzopoulos2019, Hosseinzadeh2022}, leaving both the nature of the light-curve fluctuations and the applicability of the magnetar model unresolved. 
Here, we report high-cadence multiband observations of an SLSN-I with clear ``chirped'' (i.e., decreasing period) light-curve bumps that can be directly linked to the properties of the magnetar central engine. 
Our observations are consistent with a tilted, infalling accretion disk undergoing Lense-Thirring precession around a magnetar centrally located within the expanding supernova ejecta. Our model demonstrates that the overall light curve and bump frequency independently and self-consistently constrain the spin period of the magnetar to $P=4.2\pm 0.2 \textrm{ ms}$ and the magnetic field strength to $B=1.6\pm0.1 \times10^{14}$ G. Assuming standard accretion disk parameters \citep{FKR}, we constrain the accretion rate onto the magnetar to $\dot{M}(t=30\mathrm{\ days})=2.1\pm0.19\times10^{-5} \ M_\odot \mathrm{\ yr^{-1}}$.    
Our results provide the first observational evidence of the Lense-Thirring effect in the environment of a magnetar, and confirm the magnetar spin-down model as an explanation for the extreme luminosity observed in SLSNe-I.
We anticipate this discovery will create avenues for testing general relativity in a new regime---the violent centers of young supernovae.
}




\maketitle
\vspace{-1cm}
\setcounter{tocdepth}{1}
\tableofcontents

\newpage
\section{Main}
\label{sec:main}

\begin{figure}
    \centering
    \includegraphics[width=0.91\linewidth]{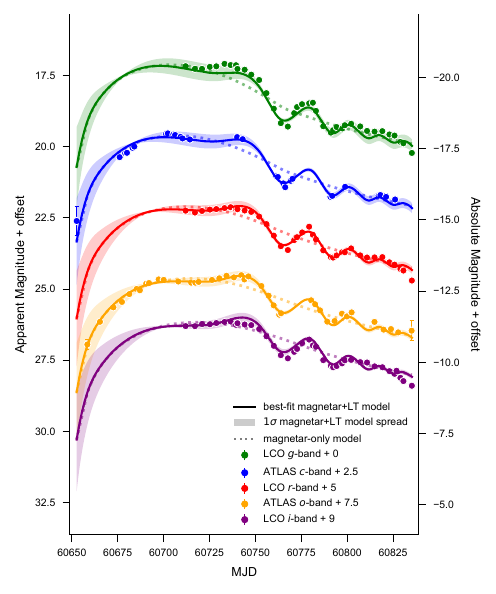}
    \caption{The combined LCO+ATLAS+KeplerCam (KC) light curves of SN 2024afav, from shortly after explosion to $\sim 125$ days post-peak. The magnetar model is shown as the dotted line (``magnetar-only''; see \autoref{sec:magnetar-only}). While the data on average are consistent with the magnetar model, there is a chirped signal post-peak that the model fails to explain. We model the light curve as generated by an infalling accretion disk undergoing Lense-Thirring precession (``magnetar+LT,'' solid line; \autoref{sec:model-construction}) and find an excellent fit to the data, with $\chi^2 \sim 1.5$ for the whole fit and $\chi^2 \sim 1.6$ for the modulations (40-180 days).}
    \label{fig:magnetar-model-fit-data}
\end{figure}

\begin{figure}
    \centering
    \includegraphics[width=1.0\linewidth]{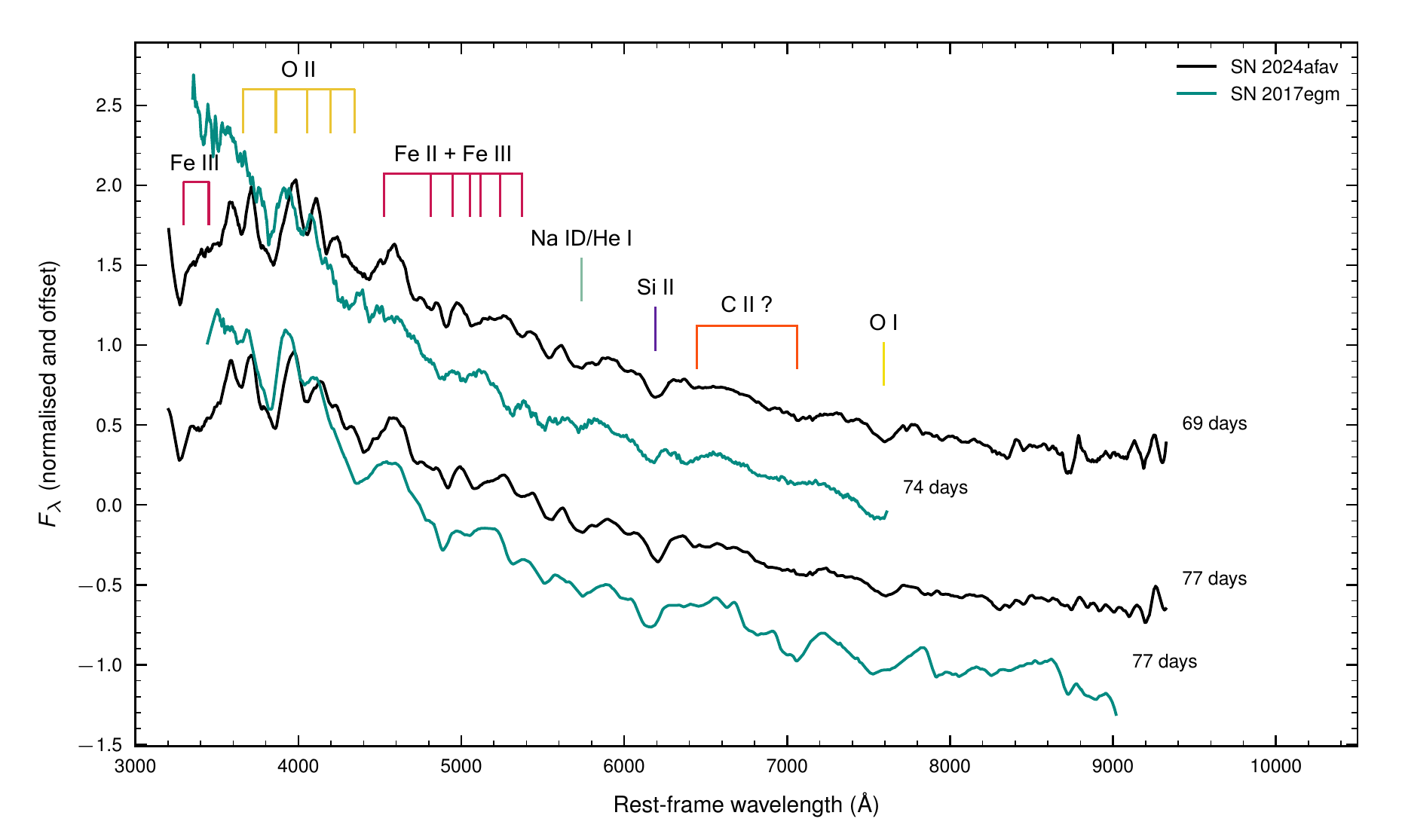}
    \caption{Comparison of the spectra of SN 2024afav (black) with the spectra of SN 2017egm (blue, \cite{Lin2023}), another well-known superluminous supernova. Lines are shown at a velocity of $\approx 8000$ km/s. The spectra of the two objects are similar, supporting the classification as a superluminous supernova, to which the magnetar model of \cite{Kasen2010} may be applicable. We note minor deviations from the spectra of SN 2017egm, such as variations in the strength of iron lines in the Fe II + Fe III complex around 5000\AA .}
    \label{fig:spectra}
\end{figure}

Supernova (SN) 2024afav is a nearby ($\sim327$ Mpc) SLSN-I which displayed the first unambiguously chirped light-curve modulations ever observed in a supernova. It was discovered by the L-GOTO Collaboration Dec 12.7, 2024 \cite{Kumar2024} and classified by ePESSTO+ as an SLSN-I on Jan 24, 2025 \citep{deWet2025} based on spectral features (see also \autoref{fig:spectra}). Early observations indicated a linear $\sim40\textrm{-day}$ rise followed by a bumpy, $\sim50\textrm{-day}$ quasiplateau phase peaking at absolute magnitude $\sim-20.5$ (\autoref{fig:magnetar-model-fit-data}). Partway through the ``plateau,'' we triggered observations using the Las Cumbres Observatory (LCO) via the Global Supernova Project and the Fred Lawrence Whipple Observatory (FLWO) KeplerCam to obtain photometry and spectra (see \autoref{sec:observations} for full discussion). Upon the appearance of a second (MJD 60736) and third (MJD 60779) sinusoidal modulation in the light curve, each with a period reduced by $\sim35\%$, we phenomenologically predicted both the epochs and luminosities of subsequent bumps and dynamically adjusted our observation campaign to capture them (see \autoref{sec:observations}). In this way, we successfully observed a fourth (MJD 60802) and fifth (MJD 60820) modulation, refining the initial period reduction fraction estimate to $\sim29\pm10\%$ (see \autoref{sec:residuals}).

In the prevailing theory to explain superluminous supernovae, a strongly magnetized millisecond pulsar (magnetar) is created in the supernova explosion, and as the spin rate decreases, it transfers some of this energy to the expanding ejecta \citep{Kasen2010,Woosley2010}.  However, while initial analytic models determined the energy budgets were sufficient to power the SLSN, they did not investigate a mechanism of energy transfer. 

The magnetar model predicts a rapid rise followed by a smooth, monotonic decline. Despite the magnetar model predicting smooth monotonic behavior in the light curve post-peak, a majority of SLSNe-I are observed to have bumps or modulations of some kind \citep{Lunnan2018,Hosseinzadeh2022}.  
Within the context of a magnetar-powered model, previous studies have attempted to explain light-curve fluctuations in SLSNe-I by invoking either aperiodic \citep[e.g., magnetar flares;][]{Dong2023} or completely periodic \citep[e.g., misaligned magnetic jet precession;][]{Zhang2025} modulations of the central engine, which have been successfully modeled against light curves with $\leq2$ bumps. The standard magnetar model fails to explain the unusual chirped signal in SN 2024afav, as shown in \autoref{fig:magnetar-model-fit-data} (see also, \autoref{fig:magnetar-only-fit}). While some number of bumps are common in SLSNe-I light curves \citep{Hosseinzadeh2022}, existing models are fundamentally incompatible with a light curve with $\geq5$ unambiguous modulations displaying a decaying period. 

\begin{figure}
    \centering
    \includegraphics[width=1.0\linewidth]{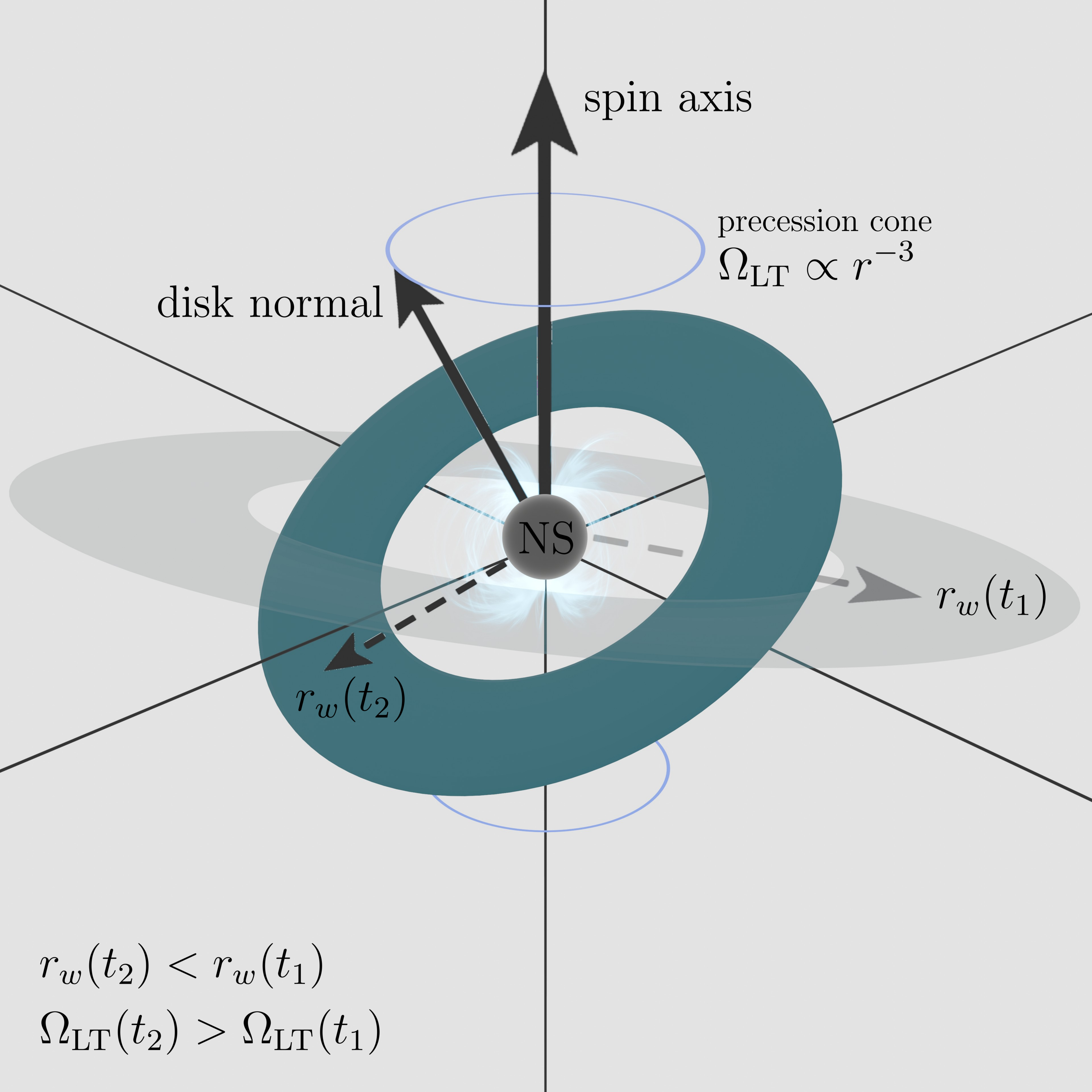}
    \caption{Schematic diagram showing the physical picture motivating the magnetar+LT model. The magnetar is shown in the center of the grid (gray sphere, NS), producing emission which is modulated along the observer's line of sight by an accretion disk (blue annulus). The misalignment between the accretion disk and the spin axis of the neutron star induces a Lense-Thirring torque that causes the disk to precess (with frequency $\Omega_{\mathrm{LT}}\propto r^{-3}$) around the spin axis, producing modulations in the light curve from the perspective of the observer as the disk periodically blocks or reflects the spin-down energy from the magnetar. As the accretion disk infalls from a larger radius $r_w(t_1)$ (gray annulus) to a smaller one $r_w(t_2)$ (blue annulus), the induced Lense-Thirring torque increases and the precession (and therefore modulation frequency) increase as well.}
    \label{fig:visual}
\end{figure}

A young magnetar is expected to form a small, tilted accretion disk from matter originating in the progenitor star that failed to eject during the supernova (\autoref{fig:visual}) \citep{Ogilvie2001,Lin2021}. The misaligned accretion disk is supported by a wind launched from the magnetar \citep{Chashkina2019,Tamilan2025}, which causes the disk to fall inward as the accretion rate in the disk decreases (\autoref{subsubsec:precession}). Any misalignment between the accretion disk and the spin axis of the neutron star would induce a Lense-Thirring torque on the accretion disk \citep[see e.g.,][for a review]{Mashhoon1984}, causing it to precess (with increasing frequency due to the infall) around the magnetar spin axis. As the accretion disk precesses, it may periodically modulate the emission from the magnetar by, for example, alternately obscuring or reflecting the magnetar luminosity \citep[as in, e.g.,][]{Jurua2011,Romanova2013}, redirecting a jet along the normal of the accretion disk \citep[as in, e.g.,][]{Liska2018}, redirecting wind from the disk itself \citep[as in, e.g.,][]{Dexter2013}, or changing the accretion rate onto the magnetar \citep{Nixon2012}. The modulating emission from the magnetar leads to an asymmetric deposition of energy in the supernova ejecta that precesses with the accretion disk and appears to an observer to oscillate \citep[e.g.,][]{Zhang2025} on the Lense-Thirring timescale of $P_{\textrm{LT}}\approx25-45$ days (\autoref{sec:model-construction}). High-energy photons from the magnetar interact with the ejecta and are reprocessed into optical photons, delaying the emission and diffusing out on a timescale of $t_d\approx15$ days during the bumps \citep[][see also \autoref{sec:model-construction}]{Rybicki1986,Sonneborn2012}. The intensity of the bumps is determined not only by the energy injection \citep{Menou2001,Dexter2013} but also the reprocessing and diffusion characteristics of the expanding ejecta \citep{Arnett1982}. However, the evolving periodicity of the modulations is solely determined by the properties of the precession (\autoref{sec:model-construction}), enabling a direct probe of strong gravity effects in the vicinity of the magnetar. 

We fit our combined magnetar and Lense-Thirring precession model to the SN 2024afav light curve (labeled ``magnetar+LT'' in \autoref{fig:magnetar-model-fit-data}). Crucially, the phase component of the modulation has only two free parameters---the initial phase of the accretion disk and the accretion rate---and is otherwise entirely determined by the parameters of the magnetar (see \autoref{sec:model-construction}). We find our model accurately reproduces the behavior of the transient during both the rise and the modulations ($\chi^2\approx1.5$ for the whole fit, $\chi^2 \approx 1.6$ for the modulations; \autoref{sec:model-construction}). Assuming standard neutron star mass ($M_{\textrm{NS}}\approx 1.4 \ M_\odot$) and size ($R_{\textrm{NS}}\approx 10^6$ cm), we can separately constrain the magnetar spin period $P$ using the observed rise to peak (spin-down energy injection constraint) and the residual behavior of the modulations (Lense-Thirring precession constraint). We find the period determined from the rise in the standard magnetar model ($P= 5.5\pm1.25$ ms, \autoref{sec:magnetar-only}) and the modulations ($P= 5.4\pm1.8$ ms, \autoref{subsubsec:precession}) are statistically self-consistent and, combined with the estimate from the full light curve (\autoref{sec:magnetar-only}), constrain the spin of the magnetar to $P= 4.2\pm0.2$ ms (\autoref{sec:model-construction}). Such self-consistency is remarkable as it suggests the modulations and overall light-curve behavior can be directly derived from the same properties of the magnetar. Similarly, combining the measurements from the early light curve and the modulations, the magnetic field strength is constrained to be $B\approx1.6\pm0.1\times10^{14}$ G (\autoref{sec:model-construction}). Assuming the accretion disk is wind-supported \citep{Chashkina2019,Tamilan2025}, we constrain the accretion rate to $\dot{M}(t=30\textrm{ days}) = 2.1\pm 0.19 \times 10^{-5} \ M_\odot/\textrm{yr}$ (\autoref{subsubsec:precession}).

\begin{figure}
    \centering
    \includegraphics[width=1.0\linewidth]{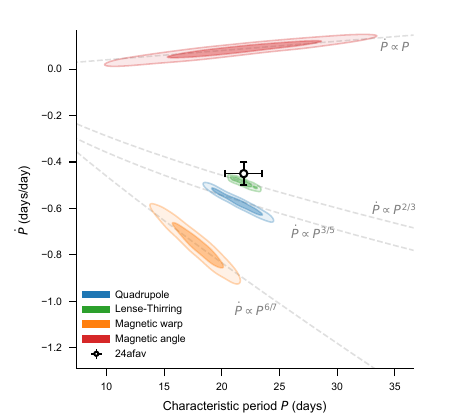}
    \caption{Ability of alternative scenarios to explain the modulation of the SN 2024afav light curve. We analyze other possible origins of the accretion disk precession to assess the uniqueness of the LT model's ability to explain the observed modulations in SN 2024afav. We characterize the candidate explanations based on their characteristic period (average period over the 80 days when the bumps are visible) and precession period derivative (see \autoref{sec:alternatives}). The corresponding location of SN 2024afav is shown as the point with error bars. $1\sigma$ and $2\sigma$ contours are shown for the candidate explanations based on the MOSFiT uncertainties (see \autoref{sec:magnetar-only}). For models not equipped with an $\dot{\Omega} > 0$ behavior, we invoked radial infall to explain the changing period. We consider magnetic angle precession, quadrupole-driven nodal precession, and magnetic warp precession. Only the Lense-Thirring effect around a magnetar produces a characteristic period and period derivative consistent with the observations. Gray dashed lines indicate the predicted $\dot{P}(P)$ behavior based on the linear radial infall assumption described in \autoref{sec:model-construction}.}
    \label{fig:alternatives}
\end{figure}

We consider alternative physical scenarios that may produce the same oscillating light-curve behavior, including various kinds of precession and input energy modulation (see \autoref{sec:alternatives}). Existing explanations that invoke a jet precessing with the magnetic inclination angle of a deformed magnetar predict approximately the correct timescale for the modulations \citep{Zhang2025} but a positive precession period derivative as the magnetar spins down, contrasting with the observations which show a decreasing period. With the assumption that a precessing accretion disk is instead modulating the line-of-sight energy injection from the magnetar into the ejecta, we consider alternative origins of the precession.
Apsidal precession is unlikely as the disk is highly circularized within minutes to hours of formation \citep{Armitage2008}. Magnetic warp precession \citep{Lai1999} and quadrupole-driven nodal precession \citep{Morsink1999,Colaiuda2008,Tremaine2014} can give comparable periodicities. However, neither can produce the required period derivative, and in the case of magnetic warp precession, the period is expected to increase owing to an induced disk-threading torque \citep{Lai1999}. By contrast, the tilted, infalling accretion disk undergoing Lense-Thirring precession gives the correct timescales and precession period derivative without deviating from the properties of the magnetar inferred from the overall behavior of the supernova (see \autoref{fig:alternatives}).

Some explanations of modulations in SLSN-I light curves have proposed interaction with circumstellar material (CSM) as the origin \citep[e.g.,][]{Liu2018,Lin2023,Hosseinzadeh2022}. In particular, the modulations in SN 2017egm were found to favor a CSM interaction origin over alternative scenarios \citep{Liu2018}. The model of \cite{Liu2018} could be used to model arbitrary modulations in an SLSN-I light curve, including SN 2024afav. However, we consider the CSM modulation mechanism less likely for several reasons. First, to reproduce $\geq5$ modulations with decaying period and amplitude requires dozens of highly fine-tuned CSM parameters. Second, the shape of the modulation is firmly sinusoidal, as is visible in the residuals (see \autoref{fig:residuals}), contrasting with the more irregular shape produced by the CSM model of \cite{Liu2018}. Third, the light curve has no modulations until well beyond the diffusion timescale ($t_d \gtrsim 37$ days), consistent with a central engine-based mechanism. Finally, our proposed description of the oscillating behavior in SN 2024afav sees its goodness-of-fit improve when a diffusion correction with no additional free parameters is incorporated (\autoref{sec:model-construction}), behavior that has been used to favor a central-engine origin over CSM interaction \citep{Hosseinzadeh2022}. Thus, interaction with CSM is unlikely to be the origin of the observed light curve modulations in SN 2024afav. However, this does not preclude CSM interactions from being present at a lower level.

The uniquely high-cadence ($\sim0.5$ day) and extended ($\sim200$ day) quality of our dataset unambiguously reveal chirped modulations in an SLSN-I light curve, behavior which is evidence of a central-engine origin. Moreover, both the overall light-curve evolution and the modulations themselves are quantitatively consistent with a magnetar-powered model, confirming its viability as an explanation for the extreme luminosity of SLSNe-I and providing evidence of a direct physical mechanism to communicate the magnetar spin-down power into the ejecta. Our analysis further demonstrates that Lense-Thirring precession provides the most plausible explanation for the chirped signal, constituting observational evidence of the Lense-Thirring effect in a magnetar and establishing the influence of general relativistic frame-dragging in a new astrophysical regime: young supernovae.

\begin{figure}
    \centering
    \includegraphics[width=0.9\linewidth]{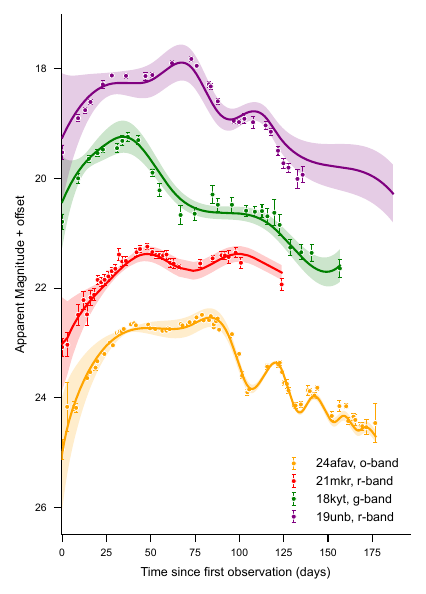}
    \caption{Fits to other SLSNe-I using the magnetar+LT model. Periodic fluctuations (with $\leq 2$ bumps observed) in SN 2018kyt (green), SN 2019unb (purple), and SN 2021mkr (red) were explained using CSM interaction, magnetic inclination precession, or central-engine flares. These solutions are either periodic on the timescale of the supernova or aperiodic, and cannot describe the chirped behavior of SN 2024afav (orange). By contrast, we find that the magnetar+LT model can explain the behavior in these other objects as well as SN 2024afav. (Note: bands for each object were selected based on which band most clearly displayed the modulations.)}
    \label{fig:comparison}
\end{figure}

We also find that our framework reproduces the behavior of previously observed SLSNe-I with periodic light-curve features (\autoref{sec:legacy}). Previous extensions of the magnetar model \citep{West2023,Dong2023,Lin2023,Zhang2025} to explain periodic behavior fail to reproduce the rapidly decaying period we observe in SN 2024afav; by contrast, our model can accommodate both chirped behavior as well as the completely periodic behavior observed in some SLSNe-I. We consider a sample of SLSNe-I which have been modeled previously using CSM interaction \citep[SN 2019unb;][]{Hosseinzadeh2022,Chen2023}, central-engine flares \citep[SN 2021mkr;][]{Dong2023}, and magnetic inclination precession \citep[SN 2018kyt;][]{Zhang2025}. We find that these objects can be satisfactorily modeled using our magnetar+LT model (\autoref{fig:comparison}). For each source, the observed period follows directly from the best-fit magnetar spin and magnetic field strength, which fix the Lense-Thirring torque and hence the precession frequency (see \autoref{sec:model-construction}, \autoref{sec:legacy}). 
Thus, SLSNe described by three different mutually inconsistent physical models can now be described by one. 

Forthcoming wide-field surveys, particularly the Legacy Survey of Space and Time (LSST) \citep{Ivezic2019,Tyson2002}, are expected to discover thousands to tens of thousands of SLSNe-I \citep{Villar2018}, which may be followed with facilities like LCO to acquire the extended observations needed to detect the fainter modulations predicted by the magnetar+LT model. This sample would provide a powerful testbed for the magnetar model and enable population-level inference of magnetar properties. Detailed follow-up campaigns will lead to robust constraints on magnetar properties, and---when combined with spectral diagnostics of accretion and ejecta---possibly even tests of general relativity using infant magnetars.

\vspace{2cm}
\noindent \textbf{Acknowledgements:}
We thank Lars Bildsten, Omer Blaes, Sunny Wong, and Jonathan Delgado for helpful discussions. J.R.F. is supported by the U.S. National Science Foundation (NSF) Graduate Research Fellowship Program under grant No. 2139319. This work makes use of data from the Las Cumbres Observatory global telescope network. The LCO group is supported by NSF grants AST-1911151 and NSF-2308113. This research was supported in part by grant NSF PHY-2309135 to the Kavli Institute for Theoretical Physics (KITP). L.J.P. is supported by a grant from the NASA Astrophysics Theory Program (ATP-80NSSC22K0725). A.V.F. acknowledges financial support from the Christopher R. Redlich Fund
and many other donors.
This work is supported by the National Science Foundation under Cooperative Agreement PHY-2019786 (The NSF AI Institute for Artificial Intelligence and Fundamental Interactions, \url{http://iaifi.org/}).

\vspace{1cm}
\noindent \textbf{Author contributions:} J.R.F. initiated the study and conceived of the mechanism, helped organize follow-up observations of the object with LCO, processed LCO photometry and spectra, performed the analysis, and led the writing of the manuscript. L.J.P. assisted with the theoretical development of the mechanism and alternatives, and contributed text to the manuscript. L.J.P., Y.Q.N., and D.A.H. contributed equally to development of the mechanism and application to observables. C.M., M.A., H.K., D.H., S.G., K.W., A.V.F., E. B., and P.L.B. provided data, assisted with interpretations, and gave feedback on the manuscript. K.A.B. developed the \texttt{lcogtsnpipe} software which was used for reductions of LCO photometry.

\vspace{1cm}
\noindent \textbf{Author information:} The authors declare no competing financial interests. Correspondence should be addressed to J.R.F. (jfarah@lco.global).

\section{Methods}
\label{sec:methods}

\subsection{Observations and data reduction}
\label{sec:observations}

SN 2024afav is located at right ascension (R.A.) 12:49:12.050 and declination (decl) $-$18:06:12.61 (J2000 epoch), in the vicinity of the faint galaxy GALEXASC J124911.85-180609.6. The unusually bright absolute magnitude ($\sim-20.5$ in the $g$ band at peak, $K$-correction of $K=-0.06$; \cite{Hogg2002}) and a hydrogen-deficient spectrum led to a classification by the extended Public ESO Spectroscopic Survey of Transient Objects (ePESSTO+) in January 2025 as an SLSN-I at redshift $z\approx0.072$ \citep{deWet2025}. Based on the redshift and assuming flat cosmological parameters ($\Omega_M = 0.286, \Omega_\lambda=0.714, H_0 = 69.6 \mathrm{\ km \ s^{-1} \ Mpc^{-1}}$), we adopt a distance of 327 Mpc to the supernova. The host galaxy was exceptionally faint in our images, and SN 2024afav is significantly offset from the center ($\approx5^"$). Using the Na I doublet method of \cite{Poznanski2012}, we constrain the extinction due to the host to be $E(B-V)_{\textrm{H}} \lesssim 0.04$ mag. However, based on the \cite{Schlafly2011} line-of-sight extinction map, we correct photometry for a Milky Way extinction coefficient of $E(B-V)_{\textrm{MW}} = 0.0527 \pm  0.0029$ mag. We review the properties of instruments used in our campaign in \autoref{sec:supp_instruments}

\begin{figure}
    \centering
    \includegraphics[width=1.0\linewidth]{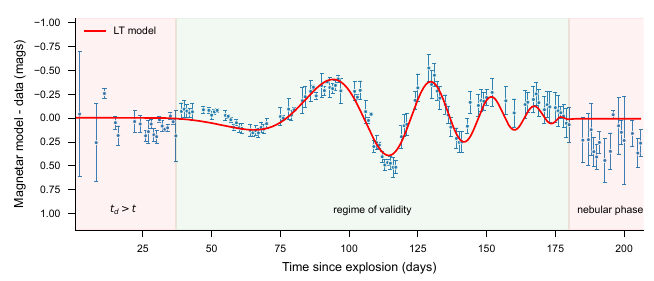}
    \caption{Visualization of the regimes of validity of the magnetar+LT model. The magnetar+LT model fundamentally assumes that the luminosity of the supernova is driven by the emission interacting with the ejecta. At early times, when the diffusion timescale $t_d > t$, modulations are not visible and thus our model is not valid. At late times, when the ejecta are optically thin (nebular phase, indicated by the spectral evolution) and no longer illuminated by the magnetar wind, our model is similarly invalid. Between these two regimes, our model may be applied, as the ejecta is sufficiently optically thick to allow high-energy photons from the magnetar to be reprocessed but sufficiently optically thin to allow the reprocessed photons to escape. }
    \label{fig:regimes}
\end{figure}

As observations were ongoing, we predicted that our model would have a regime of validity bounded on the left by the diffusion timescale $t_d$ and bounded on the right by the transition of the ejecta to the nebular phase. The magnetar+LT model assumes a luminosity driven by the magnetar wind interacting with the ejecta (see \autoref{sec:model-construction}). Once the ejecta expands to the point where it is no longer optically thick (indicated by a transition to the nebular phase, visible in e.g., the spectra), this assumption (and therefore the model) is no longer applicable. Around the time the spectra transitioned to the nebular phase, we also observed a significant deviation from luminosity power-law $L(t)\propto(1+t/t_p)^{-2}$ predicted by magnetar models (see \autoref{fig:regimes}), adding supporting evidence that the ejecta are no longer sufficiently optically thick to be illuminated by the magnetar wind. For these reasons, we constrain the regime of validity of our analysis to $37 \textrm{ days} \lesssim t \lesssim 181$ days.

\subsection{Magnetar-only model fits}
\label{sec:magnetar-only}

The unusually bright light curve ($\sim-20.5$ peak absolute magnitude) with a long rise ($>40$ days, see \autoref{sec:observations}) are consistent with predictions for a supernova explosion powered by a magnetar central engine, as described by the model of \cite{Kasen2010}. The magnetar model monotonically declines post-peak and is not expected to explain the chirped modulations. Despite this, the overall behavior of the light curve may still be used to infer parameters of the magnetar system. We fit the magnetar model of \cite{Kasen2010} to the entire dataset using the Modular Open Source Fitter for Transients \citep[MOSFiT;][]{Guillochon2018}, which has been used to fit a wide variety of SLSN-I light curves \citep[e.g.,][]{Nicholl2017,Gomez2024}. We initialize the fit with 1600 walkers and use sufficient steps ($\sim8000$, no burn-in) to achieve convergence, which we measure via chain autocorrelation length and the improved Gelman-Rubin statistic, following \cite{Farah2025a} section 5.4. Once convergence was achieved, parameter estimates and uncertainties were estimated by computing the highest posterior density interval around the maximum a posteriori value.

The results of the fit are shown in \autoref{fig:magnetar-only-fit}. All fit-parameter posteriors were unimodal with approximately symmetric uncertainties. The fit explains the rise, peak, and average decline well, but as expected, does not explain the modulations. The MOSFiT fit infers the presence of a magnetar with magnetic field strength $B= 1.62\pm0.1\times10^{14}$ G and spin period $P_{\textrm{spin}}=4.19\pm0.18$ ms, illuminating  ejecta of mass $M_{\textrm{ej}}= 10\pm1.1 \ M_\odot$ expanding with velocity ($5.754\pm 0.12)\times10^3$ km/s. The magnetar properties are consistent with expectations from theory \citep{Kasen2010} and surveys of other SLSNe-I \citep{Nicholl2017,Gomez2024}, as shown in \autoref{fig:comp_gomez}. 

We consider the possibility that the bumps may be significantly altering the outcome of the fit, as the largest modulations represent $\pm1$ mag deviations from the magnetar model. To address this, we re-perform the magnetar fit, using only data from $\lesssim50$ days (approximately the diffusion timescale). We find that fitting using this reduced sample of the data moderately affects the best-fit parameters. First, the uncertainties dramatically increase across all parameters owing to the lack of data, as only ATLAS $c$- and $o$-band data were consistently available prior to peak (\autoref{sec:observations}). The period is statistically consistent with the previous fit ($P_{\textrm{spin}}=5.49^{+0.11}_{-1.25}$ ms). The magnetic field strength and ejecta mass decrease substantially ($B=1.0\pm0.7\times10^{14}$ G and $M_{\textrm{ej}}=5.0^{+1.0}_{-4.7} \ M_\odot$, respectively). The ejecta velocity is again consistent with the previous fit and predictions from \citep{Kasen2010} at $v_{\textrm{ej}}=6.03^{+1.51}_{-2.75}\times10^3$ km/s. In both fits, the behavior of the light curve through the modulations is very similar, indicating the bumps are not significantly affecting the post-peak behavior of the model. However, the start of the modulations coincides approximately with the peak of the supernova, which is highly constraining for certain parameters (particularly the spin period, magnetic field strength, and diffusion timescale). By trimming our dataset to before the modulations began, we also limited the ability of the fit to properly constrain the peak of the light curve, possibly exacerbating discrepancies in the parameters between the two fits.

\subsection{Residual analysis}
\label{sec:residuals}

\begin{figure}
    \centering
    \includegraphics[width=1.0\linewidth]{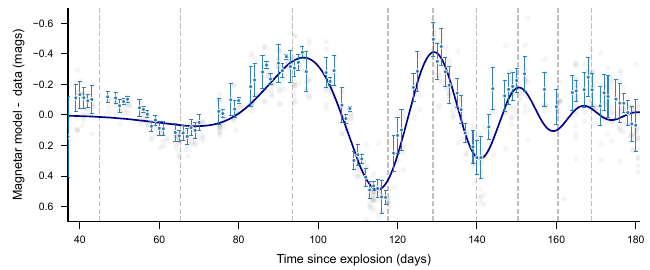}
    \caption{Residuals between the magnetar and model data across all bands phenomenologically investigated. The aggregated residuals from each band (gray points) are binned in time (blue points). We identify at least 5 clear bumps in the residual (gray dashed lines). To characterize the behavior, we fit a combined quadratic polynomial envelope and a cubic polynomial phase function, for a total of seven free parameters (dark blue line). The result of this fit indicates a hybrid growing/decaying envelope peaking around $\sim120$ days from the explosion epoch estimated by MOSFiT and a chirped signal with periodicity declining from $P\approx50$ days to $P\approx20$ days over the course of $\sim80$ days. The fit underestimates the first peak but produces peaks consistent with the remaining modulations.}
    \label{fig:poly_residuals}
\end{figure}

We characterize the modulations by computing the residual between the data and the model fit in \autoref{sec:magnetar-only}. We work in magnitude space and weight all filters equally as (i) the behavior is consistent enough in all bands to make this effective, and (ii) the spectral energy distribution (SED) of the object is quite complicated based on the spectra, which introduces uncertainty into (for example) a blackbody fit to convert to luminosity space. The effect of this assumption is to assume a flat SED, which is sufficient for our analysis but introduces minor deviations in some bands as seen in \autoref{fig:magnetar-model-fit-data}. The result of this residual and subsequent binning is shown in \autoref{fig:poly_residuals}. 
We use a standard peak-extraction algorithm \citep{find_peaks} to identify candidate extrema in each band, which we then aggregate by computing a mean location for each cluster of extrema weighted by the number of bands in which the extremum appears. Using this approach, we identify peaks at $t=65.3, 117.6, 139.9$, and $160.4$ days, and troughs at $t=44.9, 93.6, 128.9, 153.3$ days as well as a low-confidence trough detection at $t=168.8$ days. Based on the first five modulations, we estimate a precession period derivative of $\dot{P}\approx -0.44$.

Next, we investigate the sinusoidal behavior of the modulations and attempt to describe them phenomenologically. We construct a generic model to describe a sinusoid with varying periodicity and amplitude using
\begin{align}
    \Delta L(t) = \left[\sum_{a=0}^{n_a} c_a t^a\right]\cos\left[\sum_{p=0}^{n_p}c_p t^p\right],
    \label{eq:polychirp}
\end{align}
which varies the periodicity and envelope using polynomials of arbitrary degree. We find that a quadratic ($n_a=2$) polynomial envelope and a cubic ($n_p=3$) phase function is required to describe the data to $\chi^2\approx 1$. The resulting fit (visualized in \autoref{fig:poly_residuals}) does not reproduce the very first modulation but produces peaks and troughs consistent with the remaining bumps.

\subsection{Construction and fit of the Lense-Thirring modulation model}
\label{sec:model-construction}

To describe the fluctuations, we construct a physical model which can embed a chirped signal in a declining magnetar light curve. To motivate this model, we note that a period range of $\sim25-45$ days as observed (see \autoref{sec:residuals}), if manifesting from a process involving Lense-Thirring precession, corresponds to a radial scale for the best-fit magnetar of $\sim5\times10^8$ cm \citep{Mashhoon1984}. This radius is the correct order of magnitude of the magnetospheric radius \citep{FKR}, which sets the characteristic scale of an accretion disk surrounding the neutron star. If the radius of the disk decreases over time, it would increase the Lense-Thirring torque and speed up the precession significantly. 

We review the model presented in \autoref{sec:main}. In the minutes to hours following core collapse and the magnetar birth, an accretion disk composed of fallback material from the progenitor envelope forms and circularizes \citep{Ogilvie2001,FKR,Stone2012,Lin2021}. The inner radius of the disk is suspended by a wind launched from the magnetar \citep{Chashkina2019,Tamilan2025} or, in more extreme cases, the magnetosphere itself. Through some process during formation, the accretion disk becomes tilted or warped, perhaps due to the presence of a binary companion \citep{Fragile2024}, a natal kick \citep{Brandt1995}, or magnetic torques \citep{Lai1999}. The tilted disk will experience a Lense-Thirring torque, inducing a precession around the spin axis of the magnetar \citep{Mashhoon1984,Fragile2024}. As the disk precesses, it periodically modulates the injection of energy into the supernova ejecta along the line of sight, displaying an oscillating signal to a distant observer \citep{Jurua2011, Romanova2013,Zhang2025}. 

The torques induced in the disk by Lense-Thirring precession are communicated through the disk by viscous forces. In the inner disk, where Lense-Thirring torques are the strongest, viscous torques may be insufficient to transport angular momentum quickly enough to prevent the shearing of the disk into distinct rings. Indeed, various groups have found that Lense-Thirring torques can lead to disk breaking within several tens of gravitational radii \citep[e.g.,][]{Nixon2012,Raj2021,Liska2022,Musoke2023}. Ref. \cite{Nixon2012} equated the viscous and Lense-Thirring torques to show that breaking occurs within a radius $R_{\textrm{break}}$. Our best-fit parameters give an $R_{\textrm{break}}$ value comparable to the radius of the magnetar---two orders of magnitude below that of the light cylinder---indicating that the disk is unaffected by warping or breaking due to Lense-Thirring torques. A break radius comparable to the light cylinder would require a disk aspect ratio $H/R\lesssim 10^{-4}$, all else being equal, which is inconsistent with our fits of the disk properties (see \autoref{subsubsec:precession}). Thus, the Lense-Thirring torque is likely to dominate and drive the precession, and therefore the modulations in the light curve.

\subsubsection{Derivation of the precession}
\label{subsubsec:precession}
Following \autoref{sec:residuals}, we model these oscillations in a two-component model, multiplying a time-evolving envelope $A(t)$ by a sinusoid,
\begin{align}
    \Delta L_{\textrm{LT}}(t) = A(t)\cos[\phi(t)].
\end{align}
The phase function $\phi(t)$ is derived directly from the Lense-Thirring precession frequency ($\Omega_{\textrm{LT}}$) in the weak-field approximation \citep{Mashhoon1984},
\begin{align}
    \Omega_{\textrm{LT}} \approx \frac{2GJ}{c^2 r^3(t)},
\end{align}
where $J$ is the angular momentum of the magnetar and $r$ is the characteristic radius of the accretion disk. The accumulated phase $\phi(t)$ is the integral of the angular frequency with time. The exact form of $r(t)$ depends on the mechanism driving the infall; however, most can be approximated to first order as linear over the timescale of tens of days (see \autoref{fig:alternatives}). We assume $r(t)\propto t$ to first order; i.e., $r(t) = r_0 - \dot{r}(t-t_0) = \dot{r}(t_{\textrm{infall}} - t)$ where $t_{\textrm{infall}}=t_0 + r_0/\dot{r}$. Under this assumption, the phase function becomes
\begin{align}
    \phi(t) = \int \ \Omega_{\textrm{LT}}\, dt \approx -\frac{GJ}{\dot{r}^3 c^2}(t_{\textrm{infall}} - t)^{-2} - \phi_0,
    \label{eq:phasefunction}
\end{align}
where the initial phase of the disk $\phi_0$ is a free parameter, and $t_{\textrm{infall}}$ and $\dot{r}$ are determined based on the assumed infall model. Next, we assume the accretion disk is supported by the magnetar-launched wind, which causes the disk to infall as the accretion rate decreases \citep{Chashkina2019,Tamilan2025}. To approximate the equilibrium radius $r_w(t)$ where the disk pressure balances against the magnetar wind pressure, we consider the disk gas pressure \citep{FKR}
\begin{align}
    P_g \propto \dot{M}^A r^{-b},
\end{align}
where $\dot{M}\propto t^{-C}$, an approximation well-motivated for $C\approx 5/3$ at $\sim10^6$ seconds by \cite{Dexter2013}. We also consider the radiation pressure generated by the disk
\citep{FKR},
\begin{align}
    P_{r} \propto \alpha^{-0.8} \dot{M}_0^{A+0.7} t^{-C(A+0.7)} r^{-b-3/8}.
\end{align}
This combined disk gas and radiation pressure is balanced against the radiation pressure from the magnetar wind,
\begin{align}
    P_w \propto L(t)/r^2,
\end{align}
where $L(t)\propto (1+t/t_p)^{-2}$ as given by \cite{Kasen2010}, which defines $t_p$ as the spin-down timescale of the magnetar. We numerically solve for the disk radius $r_w(t)$ which satisfies $P_g + P_r \propto P_w$, 
where we adopt standard values of $A=17/20$, $b=21/8$, and $C=5/3$ from \cite{FKR}, \cite{Dexter2013}, and \cite{Menou2001}. We use this radial-infall function $r(t)$ to determine the phase function in \autoref{eq:phasefunction}.

By considering only the locations of the peaks of the modulations, we can fit the phase function directly without the rest of the light curve or describing the envelope. In this way, we can directly probe the parameters of the magnetar, independent from the rest of the light curve. We perform this fit, allowing the magnetic field strength, $\alpha$-disk parameter, accretion rate, and period to vary. We find $\alpha=0.07\pm0.03$ and constrain $\dot{M}=2.1\pm0.19\times10^{-5} \ M_\odot/$yr. Notably, we also find that the magnetic field strength is well constrained to $B=1.9\pm0.58\times10^{14}$ G, consistent with the values from the bumpless fits in \autoref{sec:magnetar-only}; similarly, we find $P_{\textrm{spin}}=5.42^{+4.53}_{-1.81}$ ms, also consistent with the bumpless fits in \autoref{sec:magnetar-only}. The resulting agreement across disjoint observables eliminates degeneracy with external power sources (CSM interaction, flares, etc.) and rules out coincidental tuning of free parameters: single $B$ and $P_{\textrm{spin}}$ values reproduce both the modulation evolution and the total luminosity evolution. Such a result unifies timing and energetics under one physical engine.

\subsubsection{Modulation reproduction with magnetar parameters}
\label{subsubsec:fit}

We combine the envelope derived in \autoref{subsubsec:envelope} and the phase function in \autoref{subsubsec:precession} to form an overall fit function,
\begin{align}
    \Delta L(t)=\left(A_0 \exp\left[{-\left(\frac{t-t_{\textrm{peak}}}{t_{\textrm{peak}}/3}\right)^2}\right]\right)\cos \left[-\frac{GJ}{\dot{r}^3 c^2}(t_{\textrm{infall}} - t)^{-2} - \phi_0\right].
\end{align}
This function has only three free parameters: $A_0$, setting the overall scale of the modulations; $\dot{M}_0$, which sets $\dot{r}$ and $t_{\textrm{infall}}$; and $\phi_0$, which is an initial condition on the angle of the disk. By assuming an infalling accretion disk periodically modulating emission from the magnetar, we have consolidated the minimum seven parameters required to describe the modulations (\autoref{sec:residuals}, \autoref{fig:poly_residuals}) into just three physically motivated parameters. In contrast to the self-consistency check performed in \autoref{subsubsec:precession}, we fix all parameters that overlap with the MOSFiT magnetar model to those values identified in \autoref{sec:magnetar-only}. Finally, we incorporate a diffusion timescale correction based on the optical depth of the expanding ejecta following \cite{Arnett1982} and \cite{Zhang2025}. This correction washes out variability at early times when the diffusion timescale $t_d \gg t$, and results in a delay at later times when $t_d \lesssim 2\pi/\Omega_{\textrm{LT}}$. The delay is absorbed into the $\phi_0$ parameter.

\begin{figure}
    \centering
    \includegraphics[width=1.0\linewidth]{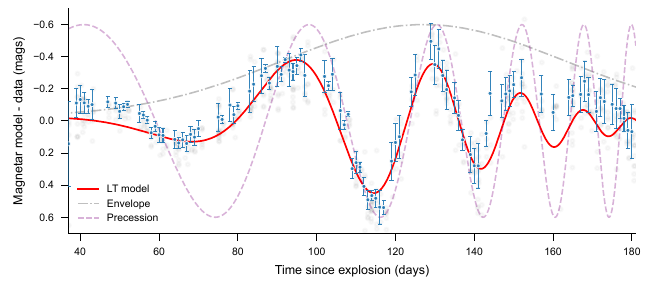}
    \caption{Same as \autoref{fig:poly_residuals}, but instead investigated using the LT model. We characterize the residual behavior using an amplitude envelope modulated by spin-down power and photon diffusion through the ejecta, combined with a precession governed by the Lense-Thirring effect. Unlike the seven-parameter phenomenological fit shown in \autoref{fig:poly_residuals}, this model only requires three parameters (overall scale, accretion rate, and initial precession angle of the disk) yet explains the data to a similar degree of success. We show the decomposition of the model into the envelope and precession, which makes the shrinking period easily visible and demonstrates the consistency between the location of the five unambiguous bumps and the prediction from the Lense-Thirring effect.}
    \label{fig:residuals}
\end{figure}

The results of the fit are shown in \autoref{fig:residuals}. Despite having fewer parameters than the phenomenological model of \autoref{sec:residuals}, the physically motivated model achieves a similar goodness-of-fit ($\chi^2 \approx 1.6$). Reincorporating the model into the overall light curve as done in \autoref{fig:magnetar-model-fit-data} emphasizes the ability of the model to explain the overall light curve and modulation behavior in conjunction with the magnetar model of \cite{Kasen2010}. We investigate the impact of the diffusion correction by attempting the fit with and without it; we find that without the diffusion correction, $\chi^2 \approx 1.9$; by contrast, the diffusion correction improves the goodness-of-fit by almost 20\% without adding any additional parameters. This behavior has been presented as evidence supporting a central-engine origin in previous works \citep[e.g.,][]{Hosseinzadeh2022}, as processes external to the ejecta would not require a diffusion correction. 

\begin{figure}
    \centering
    \includegraphics[width=1.0\linewidth]{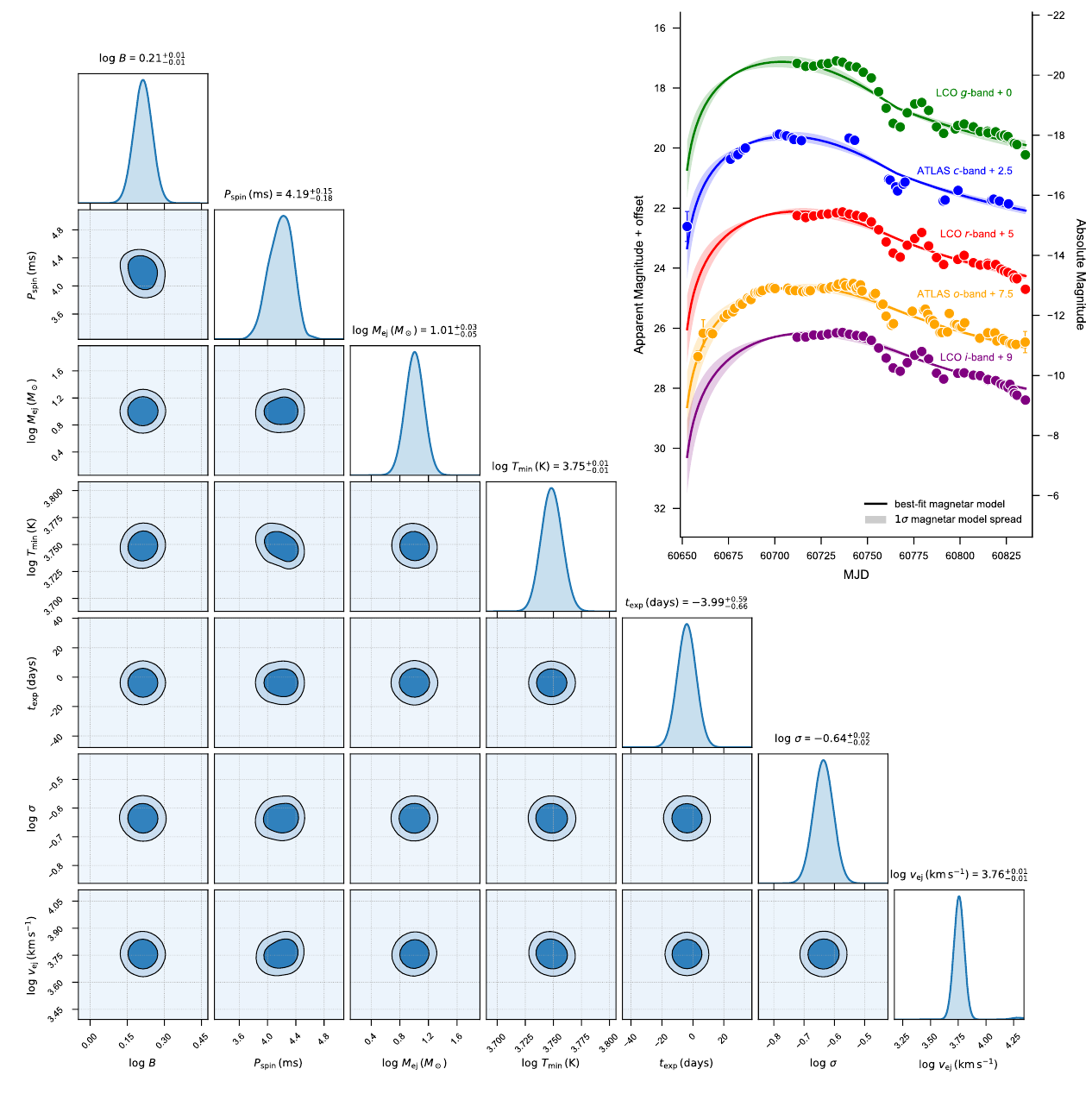}
    \caption{Corner plot and posterior samples (inset) for the magnetar-only model fit to the light curve of SN 2024afav. The posterior exploration achieved satisfactory convergence with minimal correlations between parameters and symmetric, unimodal uncertainties. The model explains the rise and overall trend of the light curve well; however, the model fails to explain the post-peak chirped modulations which appear in all bands. $1\sigma$ and $2\sigma$ regions of the joint posterior distributions are shown as contours.}
    \label{fig:magnetar-only-fit}
\end{figure}

Similarly to the polynomial evolution model of \autoref{sec:residuals}, the magnetar+LT model has difficulty reproducing the peak at $t\approx45$ days, but more successfully reproduces the trough at $t\approx 70$ days. Decomposing the model into the precession and envelope as shown in \autoref{fig:residuals}, we can immediately see that the precession model accurately reproduces the position of each peak and trough observed in the data. The inconsistency in the full model originates from the amplitude component, which despite the physical motivation is likely too simple to explain the evolution of the modulations. However, this does not affect interpretation; as demonstrated in \autoref{subsubsec:precession}, all physical quantities of interest can be estimated by fitting the extrema of our model to the dates of extrema in the light curve, without requiring a description of the amplitudes at all. There is also an inconsistency of $\approx15\textrm{-}20$ days between the location of the trough at $t\approx65$ days; however, all the other extrema are well modeled by the precession component. Incorporating estimates of the period and magnetic field from both the precession (\autoref{subsubsec:precession}), early light curve (\autoref{sec:magnetar-only}), and full light-curve evolution (\autoref{fig:magnetar-only-fit}), we estimate $P=4.2\pm0.2$ ms and $B= 1.6\pm0.1\times10^{14}$ G.

\section{Supplementary Information}
\subsection{Observing instruments}
\label{sec:supp_instruments}

SN 2024afav was almost continuously observed since explosion by the Asteroid Terrestrial Last-Alert System (ATLAS) \citep{Tonry2018}. ATLAS is comprised of dual 0.5~m telescopes located on Mauna Loa and Haleakala in Hawaii which survey the entire night sky in at least one band daily. We used the ATLAS online photometry pipeline\footnote{\href{https://fallingstar-data.com/forcedphot/}{https://fallingstar-data.com/forcedphot/}} to obtain forced photometry in the $o$ (orange) and $c$ (cyan) bands for SN 2024afav over the first $\sim180$ days of its evolution. Calibration of the ATLAS forced photometry was done to the Pan-STARRS catalog.  

Halfway through the ``plateau,'' we began observing SN 2024afav approximately daily with the Las Cumbres Observatory \citep{Brown2013} 1~m and 2~m telescopes via the Global Supernova Project, as well as obtaining roughly weekly spectra with the wide-coverage (350 nm to 1 micron) FLOYDS spectrograph located on the northern 2~m LCO instrument at Haleakala Observatory. We obtained photometry in the $g$, $r$, and $i$ bands, reduced the data using the \texttt{lcogtsnpipe} pipeline \citep{Valenti2016}, and calibrated the extracted magnitudes to AB magnitudes \citep{ABmags} using the Sloan Digital Sky Survey catalog \citep{Smith2002}. Spectroscopic data were extracted and reduced using the \texttt{floydsspec} pipeline \citep{Valenti2016}.  

We also obtained optical images of SN\,2024afav in the $gri$ filters with the KeplerCam imager on the 1.2-m telescope at FLWO. We reduced the images using standard IRAF routines, and performed photometry with the {\tt daophot} package. Instrumental magnitudes were measured by modeling the point-spread function (PSF) of each image using field stars and subtracting the model PSF from the target. To separate the flux of SN\,2024afav from that of its host galaxy we perform image subtraction on each image with {\tt HOTPANTS} \citep{Becker15}, using archival PS1/$3\pi$ images as reference templates. We estimate individual zero-points of each image by measuring the magnitudes of field stars and comparing to photometric AB magnitudes from the PS1/$3\pi$ catalog \citep{Chambers18}. The uncertainties reported for KeplerCam photometry are the combination of the photometric uncertainty and the uncertainty in the zero-point determination.

In addition to the LCO FLOYDS spectra, we also obtained spectra using the Magellan telescope in Chile and the MMT instrument at the Fred Lawrence Whipple Observatory in southern Arizona. The data were reduced using the PypeIt package \citep{PypeIt} in a standard manner. The one-dimensional spectrum was extracted and flux-calibrated using a standard star observation obtained with the same configuration.

\subsection{Derivation of the envelope}
\label{subsubsec:envelope}

We consider the modulation to the luminosity of a power source inside of an expanding ejecta which reprocesses and diffuses the photons. At any given time, the probability of emission from the magnetar being visible to the observer is scaled by the decreasing probability of the X-ray radiation from the magnetar being reprocessed by the ejecta and the simultaneously increasing probability of reprocessed photons escaping \citep{Rybicki1986,Hummer1982,Cloudy2013},
\begin{align}
    L_{\textrm{obs}}(t) \propto P_{\textrm{rep}}P_{\textrm{esc}} \propto \frac{1-e^{-\tau(t)}}{1+\tau(t)}.
\end{align}
This expression will produce a peak when $P_{\textrm{rep}}\approx P_{\textrm{esc}}$, which, given the properties of the ejecta inferred from MOSFiT imply a peak around $t\approx 300$ days, about 150 days later than is observed. However, if the input power is itself declining, as with a mass accretion rate $L\propto \dot{M} \propto t^{-5/3}$ \citep{FKR,Dexter2013}, then the expression is modified,
\begin{align}
    L_{\textrm{obs}}(t) \propto \dot{M}P_{\textrm{rep}}P_{\textrm{esc}} \propto t^{-5/3}\frac{1-e^{-\tau(t)}}{1+\tau(t)}.
    \label{eq:lum_factor}
\end{align}
The above expression has a peak at $t_{\textrm{peak}}\approx125$ days, consistent with the observed peak in the modulations, and is empirically Gaussian around the peak with standard deviation $\approx t_{\textrm{peak}}/3$. We therefore propose the following ansatz for the model envelope,
\begin{align}
    A(t) = A_0 \exp\left[{-\left(\frac{t-t_{\textrm{peak}}}{t_{\textrm{peak}}/3}\right)^2}\right],
\end{align}
where $t_{\textrm{peak}}$ corresponds to the maximum of \autoref{eq:lum_factor} and we have introduced the free parameter $A_0$, which encapsulates all initial conditions that may modify the overall modulation strength, such as (i) accretion rate, (ii) mass-energy conversion fraction $\eta$, (iii) uncertainty in the assumed distance to the supernova, (iv) inclination of the magnetar-disk system \citep[as in the model of][]{Zhang2025}, (v) spherical distribution of energy from the magnetar, etc. 

We note that previously observed SLSNe-I with periodic modulations did not always display amplitude evolution functionally similar to that of SN 2024afav. We adopt a more flexible envelope function to capture their features in \autoref{sec:legacy}.

\subsection{Investigation of alternative scenarios}
\label{sec:alternatives}

We evaluate other mechanisms that could replicate the oscillating light curve, such as different precession modes and time-variable energy injection. For each mechanism, we compute a characteristic period and average period derivative over the timescale of the modulations observed in the SN 2024afav light curve. For each mechanism, we vary model parameters to fit the characteristic period $P_c$ and calculate from those parameters a period derivative for comparison. For accretion disk precession-based mechanisms, we invoke the wind-based radial infall to produce a $\dot{P}_c < 0$ if the mechanism is not equipped with an evolving periodicity. Finally, we vary the MOSFiT magnetar model parameters and compute the corresponding $P_c$ and $\dot{P}_c$ values to generate a Monte Carlo sampling of their possible range. We compare these to the SN 2024afav values of $P_c = 23 \pm 1.6$ days and $\dot{P}_c = -0.44 \pm 0.05$. 
A visualization of the period/period-derivative parameter space of the below scenarios is shown in \autoref{fig:alternatives}. For each scenario, we compute the precession period and period derivative exactly, and compare to an analytic prediction based on assuming linear radial infall, to validate the assumption of \autoref{subsubsec:precession}. The analytic prediction is visualized for each mechanism as the gray dashed lines in \autoref{fig:alternatives}.

\vspace{0.5cm} \noindent \textbf{Magnetic inclination angle precession}: 
Ellipticity of the magnetar can lead to precession of the jet, causing modulations in the light curve \citep{Zhang2025}. We assume the model of \cite{Zhang2025} and compute the period of the magnetar versus time using their Equation 6. We adopt the MOSFiT magnetar parameters and find that a high deformation parameter of $\epsilon\approx10^{-8}$ is required to obtain $P_c \approx 23$ days. However, the combination of parameters required to reproduce the periodicity observed in the data results in $\dot{P}_c\approx0.1$, which implies an increasing periodicity. There is no mechanism in this model to create a decreasing period, and the magnitude of the predicted $|\dot{P}_c|$ is a factor of several too small. This model is disfavored.

\vspace{0.5cm} \noindent \textbf{Quadrupole-driven nodal precession}:
An oblate neutron star will cause orbits in the vicinity to precess in a manner which scales with the quadrupole coefficient of the neutron star $J_2$ and decreases with the square of the orbital radius \citep{Morsink1999}. 
There is no inbuilt mechanism for quadrupole-driven nodal precession to produce a decreasing period, so we invoke the radial infall of \autoref{sec:model-construction}. We find that an accretion rate of $\dot{M}=6.4\times10^{-6} M_\odot/\textrm{yr}$  coupled with a high $J_2 \approx 10^{-3}$ \citep{Colaiuda2008,Tremaine2014}, produces $P_c= 23\pm0.9$ days. However, the accretion rate required to reproduce the period enforces $\dot{P}_c=-0.6$; thus, this model is disfavored. 

\vspace{0.5cm} \noindent \textbf{Apsidal precession}:
If the disk has intrinsic ellipticity, it may experience apsidal precession due to relativistic effects \citep[e.g.,][]{MTW}. However, this effect is unlikely to be relevant on the timescale of the modulations observed in SN 2024afav. The disk circularizes on a timescale of roughly seconds to hours \citep{Armitage2008,Strubbe2010,Stone2012}, whereas the bumps first appear $\sim2$ months into the evolution of the supernova. By this point, the accretion disk has circularized and does not experience apsidal precession.

\vspace{0.5cm} \noindent \textbf{Magnetic warp precession}:
A diamagnetic accretion disk experiences a precessional torque with characteristic angular frequency given by \cite{Lai1999}.
This magnetic warp precession can give comparable periodicities if the disk surface density is unusually high ($\Sigma \gtrsim 10^7\mathrm{\ g \ cm^{-2}}$), but the period is again likely to increase due to an induced disk threading torque \citep{Lai1999}, conflicting with observations. If we instead modify the period by invoking radial infall, we can obtain $P_c=14.3\pm1.0$ days using an accretion rate of $\dot{M}=3.5\times10^{-5} \mathrm{\ M_\odot / yr}$. However, this results in a best-fit $\dot{P}_c=-0.69$, which is once again inconsistent with the observations of SN 2024afav; thus, this model is disfavored.

\subsection{Fits to other SLSNe-I}
\label{sec:legacy}

We consider previously studied SLSNe-I to which our model may be applicable. A majority of SLSNe-I have bumps or modulations of some kind \citep[e.g.,][]{Hosseinzadeh2022}, and some appear to be periodic, motivating oscillating physical processes to explain them \citep{Zhang2025}. A variety of physical origins are plausible for SLSNe-I modulations; we do not expect our model will be sufficient to describe all of them, especially those with non-sinusoidal or non-periodic bumps. The magnetar+LT model can be considered when the early light curve is well-described by the magnetar model and the modulations are clearly periodic or chirped in nature. We consider the following candidates which match these criteria to attempt to describe with our model.
\begin{enumerate}
    \item SN 2018kyt: discovered 2018 Dec. 19 at R.A. 12:27:56.240 decl. +56:23:35.59 by ZTF \citep{18kyt_discovery}. It was classified as an SLSN-I by ZTF based on a spectrum obtained on 2019 Jan. 29 \citep{18kyt_classification}. SN 2018kyt had two clear peaks and possibly two dips at the same cadence as the peaks ($\sim70$ days). SN 2018kyt was previously considered as a possible candidate for modulation due to magnetic inclination precession \citep{Zhang2025}.
    \item SN 2019unb: discovered 2019 Oct. 20 at R.A. 09:47:57.010 decl. +00:49:35.94 by ZTF \citep{19unb_discovery}. It was classified as an SLSN-I by ZTF based on a spectrum obtained on 2020 Feb. 22 \citep{19unb_classification}. SN 2019unb had two clear peaks and two clear dips with similar cadences ($\sim45$ days). SN 2019unb was previously considered as a possible candidate for modulation due to CSM interaction \citep{Hosseinzadeh2022,Chen2023}.
    \item SN 2021mkr: discovered 2021 May 15 at R.A. 14:56:52.570 decl. +60:50:05.10 by ZTF \citep{21mkr_discovery}. It was classified as an SLSN-I by ZTF based on a spectrum obtained on 2021 Aug. 09 \citep{21mkr_classification}. SN 2021mkr had two clear peaks and one dip with possible cadence of $\sim60$ days. SN 2021mkr was previously considered as a possible candidate for modulation due to central-engine flares \citep{Dong2023} or alternatively magnetic inclination precession \citep{Zhang2025}.
\end{enumerate}
For each object, we downloaded publicly available data produced by the Zwicky Transient Facility \citep{Bellm2019}. The ZTF survey observes on a nightly basis using a dedicated camera mounted on the 48-inch Palomar Observatory telescope. We obtained subtracted and extracted magnitudes in the $r$ and $g$  bands preprocessed by the ZTF photometry reduction pipeline \citep{Masci2023}.

To apply the model to these data, we incorporate the phase function derived in \autoref{subsubsec:precession}. However, we note from visual inspection that the candidate objects have a broader range of amplitude behaviors than SN 2024afav. To accommodate this, we replace diffusion probability with a more general exponential expression, i.e., 
\begin{align}
    A(t) = A_0 \exp\left[{-\left(\frac{t-t_{\textrm{peak}}}{\beta}\right)^2}\right],
\end{align}
where $\beta$ is a new free parameter which sets the rise and decay timescale of the envelope evolution.
Otherwise, the procedure for fitting the candidate objects is identical to our procedure for fitting SN 2024afav.

\begin{figure}
    \centering
    \includegraphics[width=1.0\linewidth]{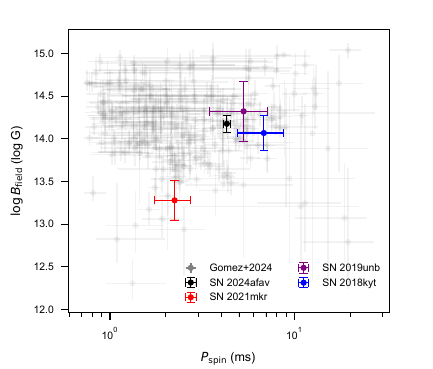}
    \caption{Comparison of $P_{\textrm{spin}}$ and $B_{\textrm{field}}$ between objects modeled with the magnetar+LT mechanism (SN 2024afav, SN 2021mkr, SN 2018kyt and SN 2019unb) and the sample of SLSNe-I provided in \citep{Gomez2024}. The parameters inferred by our model are consistent with the overall distribution of SLSNe-I.}
    \label{fig:comp_gomez}
\end{figure}

The results of the fits are shown in \autoref{fig:comparison}. We find that our model is able to satisfactorily explain the modulations in the data with goodness-of-fit in the range $\chi^2 \approx 1.3\textrm{--}1.9$, and, as shown in \autoref{fig:comp_gomez}, estimates values for $P_{\textrm{spin}}$ and $B_{\textrm{field}}$ within normal ranges for SLSNe-I \citep{Gomez2024}. The key result is that the periodicity observed in the light curve for each object is well approximated by assuming Lense-Thirring precession of an accretion disk at the equilibrium radius, given the best-fit magnetar model parameters from the bulk behavior of the light curve. Successfully fitting the magnetar+LT model to legacy objects suggests SN 2024afav may not be a fluke or unique object, and this mechanism may be relatively common in young magnetars. However, multiple SLSNe-I have been followed to hundreds of days with relatively high cadence data \citep[e.g.,][]{GalYam2009} without evidence of periodic modulations, which may have several explanations. First, the mechanism proposed here is not required to occur in every young magnetar system. Lack of fallback material or insufficient accretion rate may prevent the formation of an accretion disk. Alternatively, an insufficient natal kick (or similar) may result in the disk forming with little to no tilt, resulting in no Lense-Thirring torque. Second, the intensity of the modulations is impacted by the diffusion timescale and inclination of the system relative to the observer line of sight, so for particular system configurations (e.g., spin axis parallel to line of sight) no modulations will be visible. Additionally, our dataset is relatively unique in that, anticipating subsequent bumps, we maintained the intense cadence across all LCO bands to capture them with high S/N and on a subdaily cadence. Had we only observed the first 2--3 bumps clearly---as is the case for most SLSNe-I with modulations (see, e.g., \cite{Hosseinzadeh2022} or \cite{Zhang2025})---we would have been unable to disfavor completely periodic or aperiodic processes.

\vspace{2cm} \noindent \textbf{Data availability}: The data that support the plots within this paper are available on WiseREP.

\vspace{1cm} \noindent \textbf{Code availability}: N/A.

\section{Extended data}

\bibliography{sn-bibliography}

\end{document}